# Demonstrating Superior Discrimination of Locally Prepared States Using Nonlocal Measurements

G. J. Pryde,[1,2,*] J. L. O'Brien,[1,2,*] A. G. White,[1,2,*] and Stephen D. Bartlett[2,3]

[1]*Centre for Quantum Computer Technology, Physics Department, The University of Queensland, Brisbane 4072, Australia*
[2]*School of Physical Sciences, The University of Queensland, Brisbane 4072, Australia*
[3]*School of Physics, University of Sydney, NSW 2006, Australia*

We experimentally demonstrate the superior discrimination of separated, unentangled two-qubit correlated states using nonlocal measurements, when compared with measurements based on local operations and classical communications. When predicted theoretically, this phenomenon was dubbed "quantum nonlocality without entanglement." We characterize the performance of the nonlocal, or joint, measurement with a payoff function, for which we measure $0.72 \pm 0.02$, compared with the maximum locally achievable value of $2/3$ and the overall optimal value of $0.75$.

Quantum nonlocality, demonstrated by using entanglement to violate Bell inequalities [1,2], is one of the most profound discoveries of modern science. The nonlocal nature of entanglement is also the essential resource for many quantum information tasks including teleportation [3] and superdense coding [4]. Quantum nonlocality was thought to arise solely through the preparation of entangled states, until the recent theoretical prediction of a complementary effect in unentangled systems, which Bennett *et al.* named *quantum nonlocality without entanglement* [5–7]. In contrast to nonlocality using entanglement, where locally prepared systems exhibit nonlocal correlations when measured separately, quantum nonlocality without entanglement arises when independently prepared systems are measured jointly [8], revealing more information than can be obtained by measuring them separately (Fig. 1). This effect has practical applications, enabling increased capacity for classical communication on quantum channels [9–14] and optimally efficient quantum state estimation [15]. Here we present a demonstration of quantum nonlocality without entanglement: two photons are prepared in states that are classically correlated but *not* entangled, and a joint, or nonlocal, measurement of these two photons is shown to provide more information about the nature of the correlations than is possible using any sequence of separate local measurements.

The term "nonlocality without entanglement" should not be misunderstood: it is not that the protocol as a whole contains no entanglement, as indeed the measurement may be entangling. The key to nonlocality without entanglement is this: sets of local, possibly spacelike separated preparation events on distinct quantum systems may require a *measurement that cannot be performed locally* in order to be optimally distinguished. The fact that nonlocality needs to be introduced to optimally measure inherently unentangled, "local" systems strongly challenges typical physical intuition.

Our demonstration of quantum nonlocality without entanglement takes the form of a specific parameter estimation problem. Two parties, Alice and Bob, each prepare a single photon's polarization in a pure state, $|\psi_A\rangle$ and $|\psi_B\rangle$, respectively. Although these states are unentangled (i.e., in a product state), their preparations are classically correlated such that their states are either identical, $\langle\psi_A|\psi_B\rangle = 1$, or orthogonal, $\langle\psi_A|\psi_B\rangle = 0$. Without any prior information about these states other than that they are in a correlated product state, a third party (Charlie) wishes to perform a measurement on the pair of photons and present his best estimate of the nature of the correlation, either identical or orthogonal. We emphasize that Charlie's measurement is a one-off ("single shot") measurement; he makes his estimate of the correlation based solely on the measurement of a single pair of photons.

What makes this task difficult for Charlie is that, beyond the simple correlation of identical or orthogonal preparations, the states $|\psi_{A,B}\rangle$ are chosen randomly (i.e., from a uniform distribution on the Poincaré sphere [16]), and this choice is not known to Charlie. If these were classical systems, Charlie could simply measure each system individually, determining all of their properties and thus the

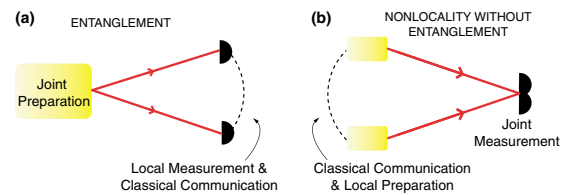

FIG. 1 (color online). Conceptual representation of (a) nonlocality *using* entanglement and (b) nonlocality *without* entanglement. The former requires joint preparation through an entangling interaction, but can be observed with local measurement and classical communication (LOCC). The latter involves separate (local) preparation, but is revealed by joint measurements, i.e., measurements not realizable with LOCC.

nature of the correlation. However, as they are quantum systems, Charlie cannot completely determine each photon's state through a single measurement. As a result of the random choice of state and the quantum nature of the systems, Charlie cannot distinguish the two types of correlations with certainty; he must perform a measurement to discriminate the possible correlations [17,18] and, based on the results of his measurement, put forward his best estimate. As we will demonstrate, Charlie has a better chance of estimating correctly if he performs joint measurements, rather than separate local ones, on the two systems even though they are unentangled: a striking example of quantum nonlocality without entanglement as originally envisaged by Peres and Wootters [5].

We allow Charlie to perform a *generalized measurement* [19], expressed as a set of positive operators $\{E_i\}$ that sum to the identity, with $i$ an index that labels the measurement outcomes. For this problem, two-outcome measurements are sufficient. We therefore restrict ourselves to two outcome measurements with $i = 0$ corresponding to "orthogonal," and $i = 1$ to "identical." That is, if Charlie performs the generalized measurement $\{E_0, E_1\}$ and gets the measurement outcome 0 or 1, he estimates that the systems were prepared in orthogonal or identical states, respectively.

To quantify Charlie's performance, we use a simple payoff function—yielding a payoff of 1 if Charlie estimates correctly and 0 otherwise—and calculate his expected payoff. Let $\rho_{\psi,j}$ be the two-qubit density matrix corresponding to the pure product state $|\psi_A\rangle \otimes |\psi_B\rangle$ where Alice and Bob have chosen orthogonal or identical preparations corresponding to $j = 0, 1$, respectively. The prior distribution is such that each $j$ occurs with probability $1/2$ and $|\psi_A\rangle$ is chosen randomly and uniformly from the Bloch sphere. If Charlie performs a generalized measurement $\{E_i\}$, the probability $p(E_i|j)$ of obtaining the measurement outcome $i$ given that the classical correlation of the preparation was $j$, averaged over all states $\psi$, is

$$p(E_i|j) = \int_S d\psi \, \text{Tr}[E_i \rho_{\psi,j}] = \text{Tr}[E_i \rho_j], \quad (1)$$

where $\rho_j = \int_S d\psi \rho_{\psi,j}$ for $j = 0, 1$, and $d\psi$ is the uniform measure on the Bloch sphere $S$. Charlie's expected payoff $\bar{P}$ using the generalized measurement $\{E_i\}$ is thus $\bar{P} = \frac{1}{2} \text{Tr}[E_0 \rho_0] + \frac{1}{2} \text{Tr}[E_1 \rho_1]$. This expected payoff $\bar{P}$ has the simple interpretation of the average probability of Charlie estimating the correlation correctly based on his measurement outcomes.

We use some symmetry properties of the states $\rho_j$ to determine Charlie's *optimal* measurement, i.e., one that maximizes his expected payoff. We note that $\rho_j$ for $j = 0, 1$ have the form of Werner states [20]

$$\rho_j = q_j \Pi_A + (1 - q_j) \Pi_S / 3,$$

where $\Pi_A$ is the projection onto the antisymmetric subspace (spanned by the singlet state $|\psi^-\rangle$ [21]) and $\Pi_S$ is the projection onto the symmetric subspace (spanned by the remaining three triplet Bell states). Note that $\rho_1$ is symmetric, and thus $q_1 = 0$, where as $\rho_0$ is neither symmetric nor antisymmetric, and is given by $q_0 = 1/2$. Thus, Charlie's estimation procedure reduces to the state discrimination problem of distinguishing the two unentangled mixed states $\rho_0 = \Pi_A/2 + \Pi_S/6$ and $\rho_1 = \Pi_S/3$. Because these two states are nonorthogonal, they cannot be discriminated with certainty.

In order to maximize his expected payoff, Charlie must perform a joint measurement on the two qubits. The expected payoff for an optimal measurement is strictly greater than the optimal local measurement [18]. We now describe both these measurements.

An optimal measurement is given by $E_0 = \Pi_A$ and $E_1 = \Pi_S$ (Ref. [18]), resulting in an expected payoff of $\bar{P}_{\text{opt}} = 3/4$. This measurement can be implemented by measuring the two qubits in the Bell state basis: if the singlet state $|\psi^-\rangle$ is detected, Charlie estimates that the states were orthogonal, whereas if any of the remaining triplet Bell states are measured then Charlie estimates that the states were identical. We emphasize that this measurement *cannot* be implemented using separate local measurements on each qubit, even if we allow classical communication.

An optimal *local* measurement is given by performing a projective measurement on each qubit along the same (arbitrary) axis and registering whether the outcomes are identical or not, which clearly requires only local operations [18]. For example, each photon could be measured in the horizontal/vertical polarization ($H, V$) basis. This measurement leads to an expected payoff of $\bar{P}_{\text{local}} = 2/3$, *less than* that of an optimal joint measurement. This result, that the optimal measurement cannot be performed locally, is remarkable because the states to be distinguished, $\rho_0$ and $\rho_1$, are both unentangled. The joint measurement therefore allows a demonstration of quantum nonlocality without entanglement.

We now describe our experimental demonstration. Using parametric down-conversion (PDC) and one-qubit rotations, we prepared correlated but *unentangled* single photon pairs from two sets of polarization input states: $|\Psi_\parallel\rangle_i \in \{|HH\rangle, |VV\rangle, |DD\rangle, |AA\rangle, |RR\rangle, |LL\rangle\}$ and $|\Psi_\perp\rangle_i \in \{|HV\rangle, |VH\rangle, |DA\rangle, |AD\rangle, |RL\rangle, |LR\rangle\}$ [22]; see Fig. 2. These sets result in the same estimation problem as that of a uniform distribution of Alice's states over the Poincaré sphere; the first with parallel ($\parallel$) correlations and the second with orthogonal ($\perp$) correlations between the single photon states. It can be readily verified that

$$\rho_\parallel = \tfrac{1}{6} \Sigma_i |\Psi_\parallel\rangle_i \langle \Psi_\parallel|_i = \Pi_S/3$$

and

$$\rho_\perp = \tfrac{1}{6} \Sigma_i |\Psi_\perp\rangle_i \langle \Psi_\perp|_i = \Pi_A/2 + \Pi_S/6$$

[23,24].

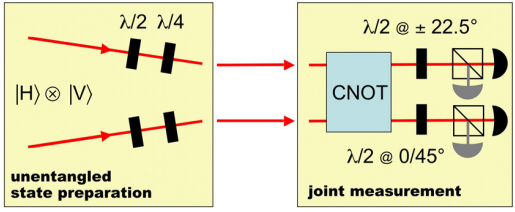
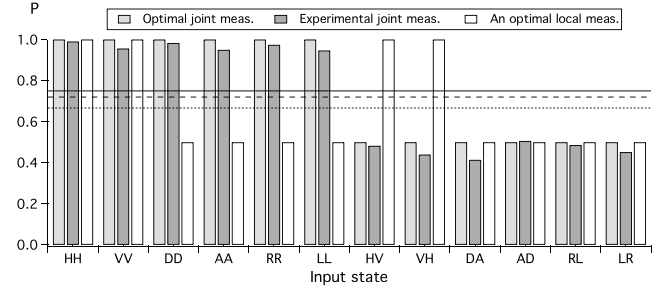

FIG. 2 (color online). Conceptual experimental setup. Pairs of degenerate unentangled photons in the state $|H\rangle \otimes |V\rangle$ are generated from a 2.5 mm type-II beta-barium borate parametric down-conversion source operating in the collapsed cone mode [31,32], pumped by an $Ar^+$ ion laser operating at 351.1 nm and 600 mW. The down-converted photons are collected into fibers to spatially filter the modes, and are prepared in identical/orthogonal product states by a half and quarter wave plate in each arm. To perform the joint measurement, we use a two photon linear optics controlled-NOT gate [25] operating with an average gate fidelity of >90%, followed by polarization analyzers, 0.36 nm FWHM interference filters, and single photon counting modules. With the output analyzers set to measure diagonal and antidiagonal in the control arm, and horizontal and vertical in the target arm, a fully resolving Bell measurement is realized whenever two photons are detected in coincidence. For our demonstration, only two detectors are used, and the probabilities of the four Bell states are built up from measurement statistics on a large ensemble of identical inputs.

To implement Charlie's optimal measurement $\{\Pi_S, \Pi_A\}$, we use a two photon controlled-NOT (CNOT) gate [25], operating with coincident detection, to measure in the Bell basis [21]. A CNOT gate takes the four maximally-entangled Bell states to four orthogonal unentangled states, which can be measured with polarizing beam splitters and detectors; see Fig. 2. This CNOT gate is nondeterministic; it does not work with unit probability, but it is known to have worked whenever one photon is measured at each of the gate's two output ports. The measurement outcome corresponding to projection onto the singlet state $|\psi^-\rangle$ is identified with $\Pi_A$, and the measurement outcomes corresponding to projection onto any of the triplet Bell states are identified with $\Pi_S$. We note that it is possible to realize a projection onto the symmetric and antisymmetric subspaces using a simple beam splitter [26], but in the absence of perfect number resolving detectors, such a scheme is also nondeterministic. Such a measurement has been used for quantum teleportation [27], as well as to estimate the overlap between two states [28]. The key advantage of employing a CNOT gate is that the probabilities for measuring the symmetric and antisymmetric outcomes are balanced (i.e., directly comparable); this is not the case for most beam splitter realizations, which require manual correction factors based on background counts.

We performed this optimal joint measurement for each of the input states in $\{|\Psi_\parallel\rangle_i\}$ and $\{|\Psi_\perp\rangle_i\}$; see Fig. 3. We calculate $\bar{P}_{exp} = 0.72 \pm 0.02$, which is unambiguously above the classical limit $\bar{P}_{local} = 2/3$: our measurement demonstrates quantum nonlocality without entanglement.

FIG. 3. Probability of correctly identifying the input state correlation (identical or orthogonal) for each of the input states used in the experiment. The bars represent a theoretical optimal joint measurement $\{\Pi_S, \Pi_A\}$; the values measured experimentally, where the CNOT is used to realize $\{\Pi_S, \Pi_A\}$; and a theoretical optimal local measurement in the $H, V$ basis [33]. The overall average payoffs, represented by horizontal lines, are 0.75, $0.72 \pm 0.02$, and $2/3$, respectively. Poissonian counting errors contribute to errors in the individual bars on the order of 0.005.

The deviation from the theoretical ideal ($\bar{P}_{opt} = 3/4$) is most likely caused by the slight optical mode mismatch in the circuit that realizes the nonlocal measurement. Perhaps the starkest contrast between what can be determined by local measurements, as opposed to joint ones, is given by the ability of our measurement to identify all parallel polarizations with nominal unit probability, $0.98 \pm 0.01$. The best local measurement achieves $2/3$ on average.

We are also able to explore some additional properties of the Bell measurement performed with a CNOT gate. One feature of this measurement is that it projects onto a *maximally entangled* basis; thus, one expects such a measurement to provide information *only* about the correlation between the two systems (in this case, classical correlations) and no information about the states of the individual systems. This result is confirmed by our data: the mutual information [19] between the four Bell measurement outcomes and the choice of Alice's preparation ($H, V, D, A, R$, or $L$) is $0.003 \pm 0.007$ bit. Another feature of the Bell measurement is that all four measurement outcomes are obtained, three of which (the triplet Bell measurement outcomes) are combined to yield the measurement $\Pi_S$. Theoretically, we expect that when the symmetric $\Pi_S$ outcome is obtained, the specific triplet Bell measurement outcomes will be completely uncorrelated with the preparation. Our data also confirms this result: the mutual information between the three triplet outcomes of the Bell measurements and the preparation by Alice and Bob is $0.0006 \pm 0.0014$ bit.

Finally, we note that although joint measurements are generally more efficient in acquiring information than local measurements, there are examples where local measurements perform equally well. Examples include discrimination between any two orthogonal, multipartite states [29] and optimal discrimination between two pure qubit states, given $N$ copies [30].

In summary, quantum nonlocality using entanglement is a defining feature of quantum physics. Here, we have demonstrated a different form of quantum nonlocality— one without entanglement—that offers insight into the structure of quantum physics. This form of quantum nonlocality, like entanglement, has important practical applications: as we have shown, joint measurements can provide more information about classical correlations between unentangled systems than is possible using local measurements. Quantum nonlocality without entanglement can offer increased performance for information processing tasks such as classical communication over a quantum channel [9–12], which has recently been demonstrated experimentally using single photons [13,14]. Also, measurements that make use of this nonlocality are required for optimal quantum state and parameter estimation and thus have applications in quantum technologies such as coherent feedback control of quantum systems.

We thank R. W. Spekkens and J. L. Dodd for stimulating discussions. This work was supported by the Australian Research Council, and by the NSA and ARDA under ARO Contract No. DAAD 19-01-1-0651.